\PassOptionsToPackage{unicode}{hyperref}
\PassOptionsToPackage{hyphens}{url}
\PassOptionsToPackage{dvipsnames,svgnames,x11names}{xcolor}
\documentclass[12pt]{article}

\usepackage{amsmath,amssymb}
\usepackage{iftex}
\ifPDFTeX
  \usepackage[T1]{fontenc}
  \usepackage[utf8]{inputenc}
  \usepackage{textcomp} 
\else 
  \usepackage{unicode-math}
  \defaultfontfeatures{Scale=MatchLowercase}
  \defaultfontfeatures[\rmfamily]{Ligatures=TeX,Scale=1}
\fi
\usepackage{lmodern}
\ifPDFTeX\else  
\fi
\IfFileExists{upquote.sty}{\usepackage{upquote}}{}
\IfFileExists{microtype.sty}{
  \usepackage[]{microtype}
  \UseMicrotypeSet[protrusion]{basicmath} 
}{}
\makeatletter
\@ifundefined{KOMAClassName}{
  \IfFileExists{parskip.sty}{%
    \usepackage{parskip}
  }{
    \setlength{\parindent}{0pt}
    \setlength{\parskip}{6pt plus 2pt minus 1pt}}
}{
  \KOMAoptions{parskip=half}}
\makeatother
\usepackage{xcolor}
\makeatletter
\ifx\paragraph\undefined\else
  \let\oldparagraph\paragraph
  \renewcommand{\paragraph}{
    \@ifstar
      \xxxParagraphStar
      \xxxParagraphNoStar
  }
  \newcommand{\xxxParagraphStar}[1]{\oldparagraph*{#1}\mbox{}}
  \newcommand{\xxxParagraphNoStar}[1]{\oldparagraph{#1}\mbox{}}
\fi
\ifx\subparagraph\undefined\else
  \let\oldsubparagraph\subparagraph
  \renewcommand{\subparagraph}{
    \@ifstar
      \xxxSubParagraphStar
      \xxxSubParagraphNoStar
  }
  \newcommand{\xxxSubParagraphStar}[1]{\oldsubparagraph*{#1}\mbox{}}
  \newcommand{\xxxSubParagraphNoStar}[1]{\oldsubparagraph{#1}\mbox{}}
\fi
\makeatother

\usepackage{longtable,booktabs,array}
\usepackage{calc} 
\usepackage{etoolbox}
\makeatletter
\patchcmd\longtable{\par}{\if@noskipsec\mbox{}\fi\par}{}{}
\makeatother
\IfFileExists{footnotehyper.sty}{\usepackage{footnotehyper}}{\usepackage{footnote}}
\makesavenoteenv{longtable}
\usepackage{graphicx}
\makeatletter
\def\maxwidth{\ifdim\Gin@nat@width>\linewidth\linewidth\else\Gin@nat@width\fi}
\def\maxheight{\ifdim\Gin@nat@height>\textheight\textheight\else\Gin@nat@height\fi}
\makeatother
\setkeys{Gin}{width=\maxwidth,height=\maxheight,keepaspectratio}
\makeatletter
\def\fps@figure{htbp}
\makeatother

\makeatletter
\@ifpackageloaded{caption}{}{\usepackage{caption}}
\AtBeginDocument{%
\ifdefined\contentsname
  \renewcommand*\contentsname{Table of contents}
\else
  \newcommand\contentsname{Table of contents}
\fi
\ifdefined\listfigurename
  \renewcommand*\listfigurename{List of Figures}
\else
  \newcommand\listfigurename{List of Figures}
\fi
\ifdefined\listtablename
  \renewcommand*\listtablename{List of Tables}
\else
  \newcommand\listtablename{List of Tables}
\fi
\ifdefined\figurename
  \renewcommand*\figurename{Figure}
\else
  \newcommand\figurename{Figure}
\fi
\ifdefined\tablename
  \renewcommand*\tablename{Table}
\else
  \newcommand\tablename{Table}
\fi
}
\@ifpackageloaded{float}{}{\usepackage{float}}
\floatstyle{ruled}
\@ifundefined{c@chapter}{\newfloat{codelisting}{h}{lop}}{\newfloat{codelisting}{h}{lop}[chapter]}
\floatname{codelisting}{Listing}

\makeatother
\makeatletter
\@ifpackageloaded{caption}{}{\usepackage{caption}}
\@ifpackageloaded{subcaption}{}{\usepackage{subcaption}}
\makeatother

\ifLuaTeX
  \usepackage{selnolig}  
\fi
\usepackage[numbers]{natbib} 

\usepackage{bookmark}

\IfFileExists{xurl.sty}{\usepackage{xurl}}{} 
\hypersetup{
  pdftitle={Title},
  pdfauthor={Author 1; Author 2},
  pdfkeywords={3 to 6 keywords, that do not appear in the title},
  colorlinks=true,
  linkcolor={blue},
  filecolor={Maroon},
  citecolor={Blue},
  urlcolor={Blue},
  pdfcreator={LaTeX via pandoc}
}

\newcommand{\anon}{0}

\usepackage{bbm} 
\usepackage{algorithm} 
\usepackage{algpseudocode} 
\usepackage{multirow} 
\usepackage{setspace} 
\newcommand{\E}{\mathbb{E}}

\newcommand{\Prob}{\mathbb{P}}
\newcommand{\iid}{\overset{\text{\footnotesize i.i.d.}}{\sim}}

\newcommand{\indep}{\perp \!\!\! \perp}
\newcommand{\one}{\mathbbm{1}}

\allowdisplaybreaks

\newtheorem{definition}{Definition}

\newtheorem{theorem}{Theorem}

\begin{document}

\def\spacingset#1{\renewcommand{\baselinestretch}%
{#1}\small\normalsize} \spacingset{1}


\if1\anon
{
  \title{\bf A Bayesian Critique of Rank-Based Methods for Surrogate Marker Evaluation}
  \author{Pietro Carlotti \\
    Department of Statistics and Data Sciences, University of Texas at Austin \\
    and \\
    Layla Parast \\
    Department of Statistics and Data Sciences, University of Texas at Austin}
  \maketitle
} \fi

\spacingset{1.8} 

\if0\anon
{
\vspace*{10mm} 
  \begin{center}
    \LARGE{A Bayesian Critique of Rank-Based Methods for \vspace*{6mm} \\
    
    Surrogate Marker Evaluation}

    \vspace*{15mm}
\normalsize Pietro Carlotti$^{1}$ and Layla Parast$^{2}$ \\
\vspace*{10mm}
\normalsize $^{1}$Department of Statistics and Data Sciences, University of Texas at Austin, \\ \normalsize 105 E 24th St D9800, Austin, TX 78705, pietro.carlotti@utexas.edu \\
\normalsize $^{2}$Department of Statistics and Data Sciences, University of Texas at Austin, \\ \normalsize 105 E 24th St D9800, Austin, TX 78705 \\

\end{center}
  \medskip
} \fi

\clearpage
\thispagestyle{empty}
\begin{abstract}

Surrogate markers are often employed in clinical trials to replace primary outcomes that may be difficult, expensive, or time-consuming to measure directly. These markers can accelerate the evaluation of new treatments, provided they reliably capture the causal relationship between treatment and true clinical benefit. Parast et al.\ \citep{parast2024rank} recently proposed a rank-based approach for evaluating surrogate markers, characterized by its nonparametric nature and minimal assumptions. While this method is useful in small-sample model-agnostic settings, it has several limitations, including a lack of clear causal interpretation, low statistical power, and insufficient robustness to different data-generating mechanisms. In this paper, we propose a Bayesian approach that addresses these shortcomings by focusing on causal treatment effect estimands and, in doing so, improves power through covariate adjustment. We demonstrate the advantages of our proposed method through a simulation study designed to highlight gains in both accuracy and power.
\end{abstract}

\noindent%
\vfill

\newpage
\setlength\parindent{15pt} 

\section{Introduction}\label{sec:introduction}

Surrogate markers are often employed in clinical trials to replace primary outcomes that may be difficult, costly, or time-consuming to measure. These markers can accelerate the evaluation of new treatments, provided they reliably capture the causal relationship between treatment and true clinical benefit. For example, Lavine et al.\ \citep{lavine2010treatment} conducted a clinical trial evaluating various treatments for children with nonalcoholic fatty liver disease, in which the primary outcome, a measure of liver function, was obtained via liver biopsy, a very invasive procedure. Instead, this study examined treatment effectiveness using a surrogate marker, change in alanine aminotransferase (ALT), a blood-based enzyme that reflects liver function, thus substantially reducing patient burden. Notably, in the United States, new treatments can be approved early based on demonstrated effectiveness on a surrogate marker via the Food and Drug Administration's (FDA) Accelerated Approval Program, established in 1992 \citep{FDA_AcceleratedApprovalProgram2024}. Many drugs have been and continue to be approved through this mechanism \citep{FDA_approved}.

A valid surrogate can substantially reduce trial burden, duration, and cost while still providing meaningful information about treatment efficacy, thereby enabling more timely access to effective treatments. However, establishing \textit{surrogate validity} is a challenging statistical problem, requiring methods that are both robust and interpretable. On this account, surrogate evaluation has been an active area of research, particularly because the FDA does not provide a formal statistical definition of a valid surrogate. A recent review by Elliott \citep{elliott2023surrogate} summarizes the state of proposed and currently used methods, including the proportion of treatment effect explained (PTE) framework \citep{freedman1992statistical,wang2002measure,parast2015robust,parast2025surrogate}, the principal stratification framework \citep{frangakis2002principal,gilbert2008evaluating}, and the meta-analytic framework \citep{buyse1998criteria,burzykowski2005evaluation,huang2011comparing}.

Unfortunately, these methods often perform poorly in small trials, especially when using robust techniques such as kernel-based estimation within the PTE framework. To address this, Parast et al.\ \citep{parast2024rank} proposed a rank-based approach for evaluating surrogate markers in such settings. Although attractive for its nonparametric nature and minimal assumptions, this method has notable limitations, including unclear causal interpretation for its target estimands, low power, and limited robustness across data-generating mechanisms.

In this paper, we improve upon this existing approach by proposing an imputation-based Bayesian alternative that extends the rank-based framework of Parast et al.\ \citep{parast2024rank}. Our approach is grounded in Bayesian causal inference, which treats unobserved potential outcomes as latent variables and samples them from their posterior predictive distributions \citep{li2023bayesian}. By using this perspective, our method not only provides a causal interpretation but also allows for covariate adjustment, improved efficiency, and a flexible framework for inference. The paper is organized as follows. In Section~\ref{sec:methods}, we review the existing rank-based approach for surrogate evaluation, outline its limitations, and emphasize how the choice of treatment effect estimands can influence conclusions about surrogate validity. In Section~\ref{sec:bayesian_approach}, we present our Bayesian methodology, including two parametric models for the data-generating process and a new criterion for the surrogate validation threshold. In Section~\ref{sec:simulation_study}, we present a simulation study comparing the performance of our proposed method to the existing approach. Finally, in Section~\ref{sec:discussion} we discuss potential future extensions.

\section{Methods}\label{sec:methods}

\subsection{Notation}\label{subsec:notation}

Let \(Y\) denote the primary outcome and \(S\) the surrogate marker. Let \(Z \in \{0,1\}\) be the treatment indicator, assigned at random, where \(Z=1\) indicates treatment and \(Z=0\) control. Without loss of generality, assume higher values of \(Y\) and \(S\) correspond to greater clinical benefit. We adopt the potential outcomes framework to define causal effects \citep{rubin1974estimating}. For each unit \(i \in \{1,\ldots,n\}\), let \(Y_{gi}\) and \(S_{gi}\) denote the potential outcomes for the primary and surrogate under treatment \(g \in \{0,1\}\). The potential outcomes for unit \(i\) are
\[
  P_{i} = (Y_{0i}, Y_{1i}, S_{0i}, S_{1i})^{\top}.
\]
We assume that for each unit \(i\), the vector of potential outcomes \(P_i\) is independently drawn from a joint distribution \(F_i\), which may differ across units, and that the treatment assignment \(Z_i\) is independently drawn from a Bernoulli distribution, with \(Z_i \indep P_i\) by design. Only the potential outcomes corresponding to the observed treatment assignment are observed. That is, each unit \(i\) contributes either (i) \( P_{1i} = (Y_{1i}, S_{1i})^{\top}\) if \(Z_{i} = 1\), or (ii) \( P_{0i} = (Y_{0i}, S_{0i})^{\top}\) if \(Z_{i} = 0\). Then, the hierarchical process that generates the data is summarized as follows:
\begin{equation}
  \label{eq:data_generating_process}
  \begin{aligned}
    (Y_{i}, S_{i})^{\top} &=
    \begin{cases}
      P_{1i}, & \text{if } Z_{i} = 1, \\
      P_{0i}, & \text{if } Z_{i} = 0,
    \end{cases} \quad i = 1, \ldots, n, \\
    P_i &\overset{\text{\footnotesize ind.}}{\sim} F_i, \quad i = 1, \ldots, n, \\
    Z_{i} &\iid \text{Bernoulli}(p), \quad i = 1, \ldots, n,
  \end{aligned}
\end{equation}
where \(p\) is the known treatment assignment probability (e.g., \(p = 0.5\) for a balanced trial).

In the next section, we describe the existing rank-based approach for evaluating a surrogate marker and call attention to its key limitations.

\subsection{Existing Frequentist Rank-Based Approach}\label{subsec:frequentist_approach}

The main idea of Parast et al.\ \citep{parast2024rank} is that if \(S\) is a valid surrogate for \(Y\), then the treatment \(Z\) should exert a comparable effect, on a probabilistic scale, on both. Their method focuses on small-sample settings, where nonparametric kernel methods are infeasible and parametric models are difficult to validate. Motivated by this, they introduced the following probabilistic index:
\begin{equation}
  \label{eq:U_definition}
  U_{Y} = \Prob(Y_{1i} > Y_{0j}) + \frac{1}{2} \; \Prob(Y_{1i} = Y_{0j}),
\end{equation}
as a measure of the treatment effect on the primary outcome \(Y\), where \(i\) and \(j\) index two different units. This quantity is often referred to as the Mann-Whitney parameter \citep{mann1947test, thas2012probabilistic, hoeffding1992class, buyse2010generalized}, as it is the estimand targeted by the corresponding hypothesis test. In this work, we restrict our attention to continuous outcomes \(Y\), thereby excluding the possibility of ties in Equation~\eqref{eq:U_definition}. Therefore, a positive treatment effect on \(Y\) is concluded if \(U_{Y} > 0.5\). The same reasoning applies to the surrogate \(S\), with \(U_{S}\) defined analogously. Then, the discrepancy in treatment effects is measured by their difference \(\delta = U_{Y} - U_{S}\). Based on this definition, Parast et al.\ \citep{parast2024rank} propose that, if \(S\) is a valid surrogate for \(Y\), \(\delta\) should be small, leading to the following hypothesis test:
\begin{equation}
  \label{eq:hypothesis_test_delta}
  \begin{cases}
    H_{0}: \delta \geq \varepsilon, \\
    H_{1}: \delta < \varepsilon,
  \end{cases}
\end{equation}
where \(\varepsilon\) is a pre-specified threshold for surrogate validation.

To estimate the parameter of interest \(\delta\), Parast et al.\ \citep{parast2024rank} compute \(\widehat{U}_{Y}\) and \(\widehat{U}_{S}\) using plug-in estimators, commonly known as the Mann--Whitney U-statistics:
\begin{equation}
  \label{eq:U_estimator}
  \begin{aligned}
    \widehat{U}_{Y} & = \frac{1}{n_{1} n_{0}} \sum^{n_{1}}_{i = 1} \sum^{n_{0}}_{j = 1} \one(Y_{i} > Y_{j}), \\
    \widehat{U}_{S} & = \frac{1}{n_{1} n_{0}} \sum^{n_{1}}_{i = 1} \sum^{n_{0}}_{j = 1} \one(S_{i} > S_{j}),
  \end{aligned}
\end{equation}
where \(n_{1}\) and \(n_{0}\) are the treated and control sample sizes, and then estimate \(\delta\) by \(\widehat{\delta} = \widehat{U}_{Y} - \widehat{U}_{S}\). A confidence interval follows from the central limit theorem for U-statistics and closed-form variance estimators for correlated U-statistics. Parast et al.\ \citep{parast2024rank} conclude that \(S\) is valid if the upper bound of this interval is less than the pre-specified threshold \(\varepsilon\).

\subsection{Limitations of the Existing Approach}\label{subsec:limitations}

The nonparametric approach of Parast et al.\ \citep{parast2024rank} provides a method for surrogate evaluation in small-sample settings, but it suffers from several notable limitations. First, it is worth questioning whether the treatment effect quantities considered truly capture the causal targets. In fact, in the causal inference literature rank-based statistics have been criticized for lacking clear causal interpretation \citep{imbens2015causal}. For example, the Mann--Whitney test assesses whether two variables are equal in distribution, but this condition is not necessarily informative about causal effects. To illustrate this point, consider that since \(U_{Y}\) represents the probability that a randomly selected treated unit has a higher \(Y\) than an independently selected control, it measures global distributional differences between groups. But is this the parameter we truly want to characterize? A more compelling target may be
\begin{align}
  V_{Y} = \Prob(Y_{1i} > Y_{0i}),
\end{align}
which represents the probability that a randomly selected unit has a better outcome \(Y\) under treatment than under control. If we define \(V_{S}\) analogously for the surrogate \(S\), then we can measure the discrepancy in treatment effects via \(\theta = V_{Y} - V_{S}\). Notably, \(U_{Y}\) and \(U_{S}\) are not equivalent to \(V_{Y}\) and \(V_{S}\), as discussed in detail by Fay et al.\ \citep{fay2018causal}. This difference propagates directly to the discrepancy parameters \(\delta\) and \(\theta\). As we illustrate in Section~\ref{subsec:comparison_parameters}, they can differ substantially even under relatively simple and well-behaved data-generating mechanisms, and therefore the choice between them may lead to opposing conclusions.

Secondly, the assumptions underlying the Mann--Whitney statistic may be too restrictive. In particular, the test assumes potential outcomes for each observation are independent and identically distributed. The identical distribution assumption is often unrealistic, as potential outcomes are typically influenced by covariates that vary across units.

Moreover, the numerical studies in Parast et al.\ \citep{parast2024rank} show that, while their estimation approach achieves minimal bias across various settings, the resulting test evaluating the surrogate suffers from low power. Approaches that prioritize higher power, even if they require some parametric assumptions, may better detect valid surrogates, especially in small-sample settings. Along these lines, leveraging baseline covariates is a natural strategy to increase power, as covariate adjustment is known to improve both power and efficiency in other settings by reducing residual outcome variability unrelated to treatment \citep{zhang2008improving, colantuoni2015leveraging, zhang2019machine}. However, the rank-based framework of Parast et al.\ \citep{parast2024rank} does not easily allow for covariate adjustment. Incorporating baseline information would likely require a rank regression--type approach, representing a substantial methodological departure from the original framework.

Hence, in Section~\ref{sec:bayesian_approach}, we describe our Bayesian imputation-based methodology addressing these limitations. In particular, our approach focuses on the causal treatment effect quantities \(V_{Y}\) and \(V_{S}\), improves power, and allows incorporation of baseline covariates. Before doing so, in the following section, we detail the difference between the discrepancy parameters; this section may be skipped without loss of continuity by readers less interested in these subtleties.

\subsection{Comparison between Discrepancy Parameters}\label{subsec:comparison_parameters}

As noted in Section~\ref{subsec:limitations}, the parameters \(U_{Y}\) and \(V_{Y}\) are not equivalent, and this difference propagates directly to the discrepancy parameters \(\delta\) and \(\theta\). In practice, this means that, under the same data-generating mechanism, the two discrepancy parameters can take substantially different values, potentially leading to opposing conclusions about surrogate validity. To illustrate this point, it is useful to consider the analytical forms of \(\delta\) and \(\theta\) in a simple setting where the potential outcomes are drawn from a Gaussian distribution with parameters that depend on a baseline covariate \(X\):
\begin{equation}
  \label{eq:data_generating_process_covariates}
  \begin{aligned}
    P_{i} &\overset{\text{\footnotesize ind.}}{\sim}\;
    \mathcal{N}_{4}\!\left( \mu^{X_{i}} ,\; \Sigma^{X_{i}} \right), \quad i=1,\ldots,n, \\
    X_{i} &\iid f, \quad i = 1, \ldots, n,
  \end{aligned}
\end{equation}
where \(\mu^{X_{i}}\) and \(\Sigma^{X_{i}}\) are the mean vector and covariance matrix that depend on \(X_{i}\), and \(f\) is the distribution of the covariate over some space \(\mathcal{X}\). Under this model, it is easy to see that
\begin{align}
    Y_{1i} - Y_{0i} \mid X_{i} = x \overset{\text{\footnotesize ind.}}{\sim} \mathcal{N} \left( \frac{ \mu^{x}_{1} - \mu^{x}_{3}}{\sqrt{\Sigma^{x}_{11} + \Sigma^{x}_{33} - 2 \Sigma^{x}_{13}}} \right), \quad i = 1, \ldots, n, \; \forall \; x \in \mathcal{X}.
\end{align}
Therefore, we have that
\begin{align}
    \begin{split}
        V^{x}_{Y} & = \Prob \left( Y_{1i} > Y_{0i} \mid X_{i} = x \right) \\
        & = \Prob \left( Y_{1i} - Y_{0i} > 0 \mid X_{i} = x \right) \\
        & = \Prob \left( \frac{Y_{1i} - Y_{0i} - \mu^{x}_{1} + \mu^{x}_{3}}{\sqrt{\Sigma^{x}_{11} + \Sigma^{x}_{33} - 2 \Sigma^{x}_{13}}} > \frac{- \mu^{x}_{1} + \mu^{x}_{3}}{\sqrt{\Sigma^{x}_{11} + \Sigma^{x}_{33} - 2 \Sigma^{x}_{13}}} \mid X_{i} = x \right) \\
        & = \Prob \left( Z > \frac{- \mu^{x}_{1} + \mu^{x}_{3}}{\sqrt{\Sigma^{x}_{11} + \Sigma^{x}_{33} - 2 \Sigma^{x}_{13}}} \right) \\
        & = 1 - \Phi \left( \frac{- \mu^{x}_{1} + \mu^{x}_{3}}{\sqrt{\Sigma^{x}_{11} + \Sigma^{x}_{33} - 2 \Sigma^{x}_{13}}} \right) \\
        & = \Phi \left( \frac{ \mu^{x}_{1} - \mu^{x}_{3}}{\sqrt{\Sigma^{x}_{11} + \Sigma^{x}_{33} - 2 \Sigma^{x}_{13}}} \right), \; \forall \; x \in \mathcal{X},
    \end{split}
\end{align}
where \(Z \sim \mathcal{N}(0,1)\) and \(\Phi\) is the standard normal cumulative distribution function. Then, by the law of total probability, we can expand \(V_{Y}\) as follows:
\begin{align}
    \begin{split}
        V_{Y} & = \Prob \left( Y_{1i} > Y_{0i} \right) \\
        & = \int_{\mathcal{X}} V^{x}_{Y} \, f (x) \, \text{d}x \\
        & = \int_{\mathcal{X}} \Phi \left( \frac{ \mu^{x}_{1} - \mu^{x}_{3}}{\sqrt{\Sigma^{x}_{11} + \Sigma^{x}_{33} - 2 \Sigma^{x}_{13}}} \right) \, f (x) \, \text{d}x.
    \end{split}
\end{align}
Since the same reasoning applies to \(V_{S}\), we can express \(\theta\) as follows:
\begin{align}
\label{eq:decomposition_theta}
    \begin{split}
        \theta & = \int_{\mathcal{X}} \theta^{x} \, f (x) \, \text{d}x \\
        & = \int_{\mathcal{X}} \left( \Phi \left( \frac{ \mu^{x}_{1} - \mu^{x}_{3}}{\sqrt{\Sigma^{x}_{11} + \Sigma^{x}_{33} - 2 \Sigma^{x}_{13}}} \right) - \Phi \left( \frac{ \mu^{x}_{2} - \mu^{x}_{4}}{\sqrt{\Sigma^{x}_{22} + \Sigma^{x}_{44} - 2 \Sigma^{x}_{24}}} \right) \right) \, f (x) \, \text{d}x,
    \end{split}
\end{align}
where \( \theta^{x} = V^{x}_{Y} - V^{x}_{S} \; \forall \; x \in \mathcal{X}\). Instead, to derive an analytical expression for \(\delta\), we can start by noting that since
\begin{align}
    Y_{1i} - Y_{0j} \mid X_{i} = x_{i}, X_{j} = x_{j} \sim \mathcal{N} \left( \frac{\mu^{x_{i}}_{1} - \mu^{x_{j}}_{3}}{\sqrt{\Sigma^{x_{i}}_{11} + \Sigma^{x_{j}}_{33}}} \right), \quad \forall \; x_{i}, x_{j} \in \mathcal{X}, \; i,j = 1, \ldots, n,
\end{align}
then we have that
\begin{align}
    \begin{split}
        U^{x_{i} x_{j}}_{Y} & = \Prob \left( Y_{1i} > Y_{0j} \mid X_{i} = x_{i}, X_{j} = x_{j} \right) \\
        & = \Prob \left( \frac{Y_{1i} - Y_{0j} - \mu^{x_{i}}_{1} + \mu^{x_{j}}_{3}}{\sqrt{\Sigma^{x_{i}}_{11} + \Sigma^{x_{j}}_{33}}} > \frac{- \mu^{x_{i}}_{1} + \mu^{x_{j}}_{3}}{\sqrt{\Sigma^{x_{i}}_{11} + \Sigma^{x_{j}}_{33}}} \mid X_{i} = x_{i}, X_{j} = x_{j} \right) \\
        & = \Prob \left( Z > \frac{- \mu^{x_{i}}_{1} + \mu^{x_{j}}_{3}}{\sqrt{\Sigma^{x_{i}}_{11} + \Sigma^{x_{j}}_{33}}} \right) \\
        & = 1 - \Phi \left( \frac{- \mu^{x_{i}}_{1} + \mu^{x_{j}}_{3}}{\sqrt{\Sigma^{x_{i}}_{11} + \Sigma^{x_{j}}_{33}}} \right) \\
        & = \Phi \left( \frac{\mu^{x_{i}}_{1} - \mu^{x_{j}}_{3}}{\sqrt{\Sigma^{x_{i}}_{11} + \Sigma^{x_{j}}_{33}}} \right) , \quad \forall \; x_{i}, x_{j} \in \mathcal{X}, \; i,j = 1, \ldots, n,
    \end{split}
\end{align}
Then, by the law of total probability, we can expand \(U_{Y}\) as follows:
\begin{align}
    \begin{split}
        U_{Y} & = \Prob \left( Y_{1i} > Y_{0j} \right) \\
        & = \int_{\mathcal{X}} \int_{\mathcal{X}} U^{x_{i}, x_{j}}_{Y} \, f (x_{i}) \, f (x_{j}) \, \text{d}x_{i} \, \text{d}x_{j} \\
        & = \int_{\mathcal{X}} \int_{\mathcal{X}} \Phi \left( \frac{\mu^{x_{i}}_{1} - \mu^{x_{j}}_{3}}{\sqrt{\Sigma^{x_{i}}_{11} + \Sigma^{x_{j}}_{33}}} \right) \, f (x_{i}) \, f (x_{j}) \, \text{d}x_{i} \, \text{d}x_{j}.
    \end{split}
\end{align}
Since the same reasoning applies to \(U_{S}\), we can express \(\delta\) as follows:
\begin{align}
    \begin{split}
        \delta & = \int_{\mathcal{X}} \int_{\mathcal{X}} \delta^{x_{i} x_{j}} \, f (x_{i}) \, f (x_{j}) \, \text{d}x_{i} \, \text{d}x_{j} \\
        & = \int_{\mathcal{X}} \int_{\mathcal{X}} \left( \Phi \left( \frac{\mu^{x_{i}}_{1} - \mu^{x_{j}}_{3}}{\sqrt{\Sigma^{x_{i}}_{11} + \Sigma^{x_{j}}_{33}}} \right) - \Phi \left( \frac{\mu^{x_{i}}_{2} - \mu^{x_{j}}_{4}}{\sqrt{\Sigma^{x_{i}}_{22} + \Sigma^{x_{j}}_{44}}} \right) \right) \, f (x_{i}) \, f (x_{j}) \, \text{d}x_{i} \, \text{d}x_{j},
    \end{split}
\end{align}
where \( \delta^{x_{i} x_{j}} = U^{x_{i} x_{j}}_{Y} - U^{x_{i} x_{j}}_{S} \; \forall \; x_{i}, x_{j} \in \mathcal{X}, \; i,j = 1, \ldots, n\).

\begin{figure}[ht]
    \centering
    \includegraphics[width=0.8\textwidth]{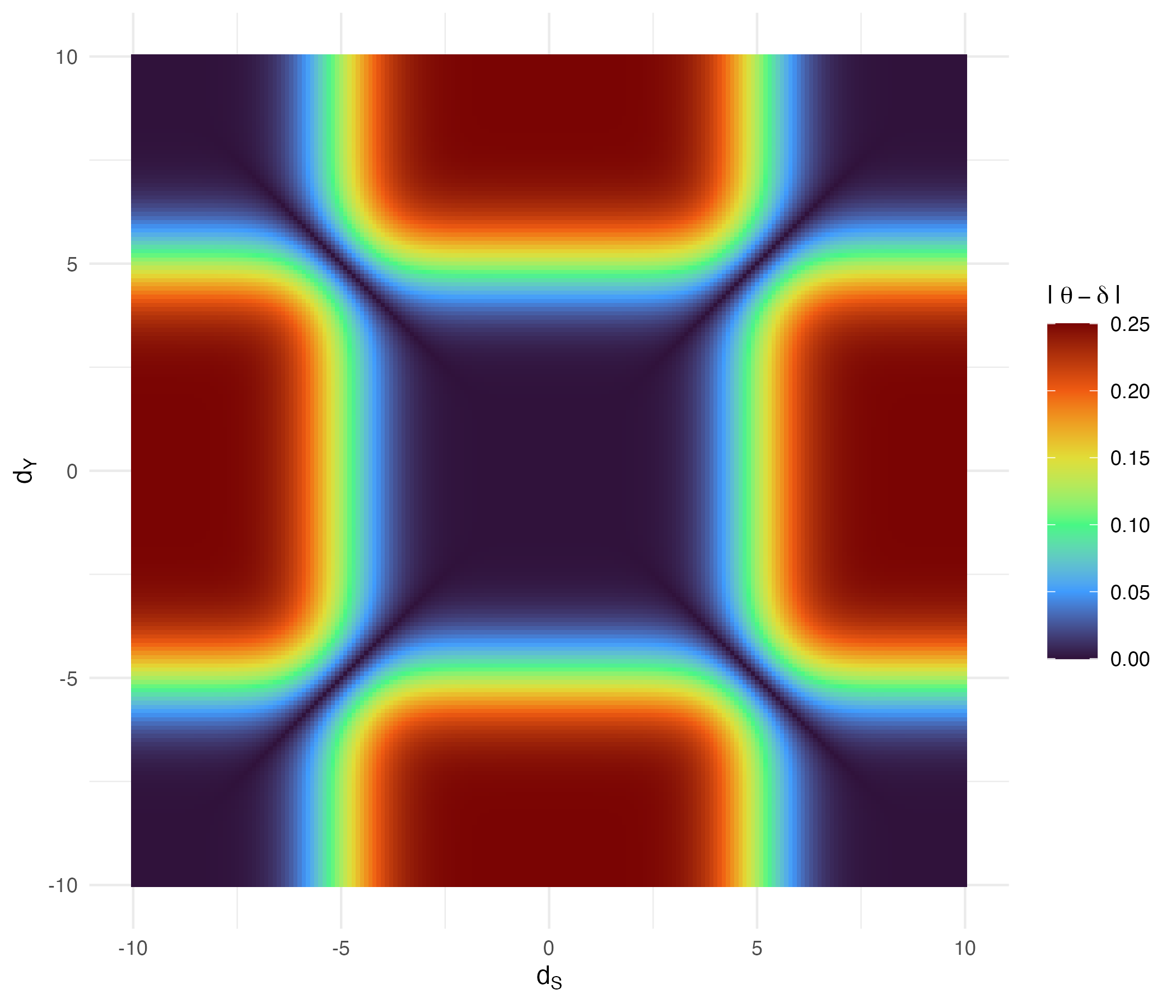}
    \caption{Heatmap of \(|\theta - \delta|\) as a function of \(d_S\) and \(d_Y\) with \(\Delta = 5\).}
    \label{fig:discrepancy_heatmap}
\end{figure}

While it may appear that the formulas for \(\delta\) and \(\theta\) are similar, they are not exactly the same. If we are willing to assume that \( \Sigma^{x}_{13} = \Sigma^{x}_{24} = 0 \; \forall \; x \in \mathcal{X} \) then it follows that
\begin{align}
  V^{x}_{Y} = U^{x x}_{Y} \text{ and } V^{x}_{S} = U^{x x}_{S} \; \forall \; x \in \mathcal{X}.
\end{align}
Moreover, if we define the function
\begin{align}
    g(x_{i}, x_{j}) = \Phi \left( \frac{\mu^{x_{i}}_{1} - \mu^{x_{j}}_{3}}{\sqrt{\Sigma^{x_{i}}_{11} + \Sigma^{x_{j}}_{33}}} \right) - \Phi \left( \frac{\mu^{x_{i}}_{2} - \mu^{x_{j}}_{4}}{\sqrt{\Sigma^{x_{i}}_{22} + \Sigma^{x_{j}}_{44}}} \right),
\end{align}
then it is easy to see that we can rewrite the two parameters as follows:
\begin{align}
    \begin{split}
        \delta & = \E_{(X_{i}, X_{j}) \sim f^{2}} \left( g(X_{i}, X_{j})\right) = \E_{X_{i} \sim f} \left( \E_{X_{j} \sim f} \left( g(X_{i}, X_{j}) \right) \right) \\
        \theta & = \E_{X_{i} \sim f} \left( g(X_{i}, X_{i}) \right),
    \end{split}
\end{align}
which means that \(\theta \neq \delta\) whenever \(\E_{X_{j} \sim f} \left( g(X_{i}, X_{j}) \right) \) and \(g(X_{i}, X_{i})\) differ in expectation. This is a quite general formulation of the issue. Therefore, to better understand when problems arise, it might be more useful to reconsider the question by imposing some simplifying assumptions. To underscore the effect of \(\mu^{x}\), suppose the potential outcomes are normalized in such a way that \(\Sigma^{x}_{11} = \Sigma^{x}_{22} = \Sigma^{x}_{33} = \Sigma^{x}_{44} = \frac{1}{2} \; \forall \; x \in \mathcal{X} \), thus simplifying \(g\) to
\begin{align}
    g(x_{i}, x_{j}) = \Phi \left( \mu^{x_{i}}_{1} - \mu^{x_{j}}_{3} \right) - \Phi \left( \mu^{x_{i}}_{2} - \mu^{x_{j}}_{4} \right)
\end{align}
Even in this very restrictive scenario, \(\E_{X_{j} \sim f} \left( g(X_{i}, X_{j}) \right) \) and \(g(X_{i}, X_{i})\) may differ in expectation even under very simple distributions \(f\) for the baseline covariate. To see this, consider a setting where \(f = \text{Bernoulli} \left( \frac{1}{2} \right) \). Although this choice may seem trivial, it can still have a meaningful interpretation in a clinical trial setting (e.g. \(X\) represents smoker vs non-smoker status). Then, we can derive a simple closed-form formula for \(\delta\) and \(\theta\):
\begin{align}
    \begin{split}
        \theta = \frac{\theta^{0} + \theta^{1}}{2}, \quad \delta = \frac{\delta^{00} + \delta^{11} + \delta^{01} + \delta^{10}}{4},
    \end{split}
\end{align}
which implies that
\begin{align}
    | \theta - \delta | = \frac{ \left| \delta^{00} + \delta^{11} - \delta^{01} - \delta^{10} \right| }{4}.
\end{align}
This expression highlights that the discrepancy between the two parameters is driven by differences between comparisons within the same covariate group and comparisons across different covariate groups. Importantly, this effect does not arise from treatment effect heterogeneity but from differences in mean potential outcomes across covariate groups. To make this more concrete, consider an even simpler setting where \(S\) is a perfect surrogate for \(Y\) and the treatment effect is homogeneous across all values of \(X\). Specifically, suppose that
\begin{align}
    \begin{split}
        \mu^{0}_{1} - \mu^{0}_{3} =
        \mu^{0}_{2} - \mu^{0}_{4} =
        \mu^{1}_{1} - \mu^{1}_{3} =
        \mu^{1}_{2} - \mu^{1}_{4} = \Delta,
    \end{split}
\end{align}
for some constant \(\Delta > 0\). Then, in this case, it follows that
\begin{align}
    \begin{split}
        \delta^{00} & = \Phi(\Delta) - \Phi(\Delta) = 0 \\
        \delta^{11} & = \Phi(\Delta) - \Phi(\Delta) = 0 \\
        \delta^{01} & = \Phi(\Delta + \mu^{0}_{3} - \mu^{1}_{3}) - \Phi(\Delta + \mu^{0}_{4} - \mu^{1}_{4}) \\
        \delta^{10} & = \Phi(\Delta + \mu^{1}_{3} - \mu^{0}_{3}) - \Phi(\Delta + \mu^{1}_{4} - \mu^{0}_{4})
    \end{split}
\end{align}
This shows that the discrepancy between \(\theta\) and \(\delta\) is driven by \(\Delta\) and the difference between means of \(Y\) and \(S\) for control units across different \(X\) groups. This implies that
\begin{align}
    |\theta - \delta| = \frac{1}{4} | \Phi(\Delta + d_{Y}) - \Phi(\Delta + d_{S}) + \Phi(\Delta - d_{Y}) - \Phi(\Delta - d_{S}) |,
\end{align}
where \(d_{Y} = \mu^{0}_{3} - \mu^{1}_{3}\) and \(d_{S} = \mu^{0}_{4} - \mu^{1}_{4}\). Then, to understand how the discrepancy between \(\theta\) and \(\delta\) changes as a function of \(d_{Y}\) and \(d_{S}\), it is useful to consider the plot of this quantity in Figure~\ref{fig:discrepancy_heatmap} for a fixed value of \(\Delta = 5\). It turns out that \(|\theta - \delta|\) is very large whenever
\begin{align}
    |d_{Y}| < \Delta \; \text{and} \; |d_{S}| > \Delta \quad \text{or} \quad |d_{Y}| > \Delta \; \text{and} \; |d_{S}| < \Delta
\end{align}
and the supremum of \(|\theta - \delta|\) is \(\frac{1}{4}\), which is achieved if \(d_{Y} = 0\) and \( | d_{S} | \to +\infty\), or vice versa. 

To understand the intuition behind this result, consider the setting where \(d_{Y} = 0\) and \(d_{S} >> \Delta\). In this case, we have that
\begin{itemize}
    \item \(U^{00}_{Y} = \Phi (\Delta)\) and \(U^{00}_{S} = \Phi (\Delta)\), therefore \(\delta^{00} = \Phi (\Delta) - \Phi (\Delta) = 0\),
    \item \(U^{11}_{Y} = \Phi (\Delta)\) and \(U^{11}_{S} = \Phi (\Delta)\), therefore \(\delta^{11} = \Phi (\Delta) - \Phi (\Delta) = 0\),
    \item \(U^{01}_{Y} = \Phi (\Delta)\) and \(U^{01}_{S} \approx 1\), therefore \(\delta^{01} \approx \Phi (\Delta) - 1 \),
    \item \(U^{10}_{Y} = \Phi (\Delta)\) and \(U^{10}_{S} \approx 0\), therefore \(\delta^{10} \approx \Phi (\Delta) - 0 = \Phi (\Delta)\).
\end{itemize}
Hence, \(|\theta - \delta| = \frac{1}{4} | 1 - 2 \Phi (\Delta)| \approx \frac{1}{4}\) for \(\Delta\) large enough (e.g., \(\Delta > 3\)) and, in particular, \(\theta = 0\) while \(\delta \approx 0.25\). This means that in this setting, even though \(S\) is a perfect surrogate for \(Y\), \(\delta\) is substantially different from \(0\), while \(\theta\) correctly captures the surrogate validity. 

The above discussion shows that the choice of discrepancy parameter is crucial, as different choices can lead to very different conclusions about surrogate validity. The rank-based approach focuses on \(\delta\), which may not capture the causal relationship between \(Y\) and \(S\) even in simple settings. We therefore advocate focusing on \(\theta\) instead.

\section{Proposed Bayesian Imputation-Based Approach}\label{sec:bayesian_approach}
The central idea of our approach is to estimate the parameter of interest and perform inference through its posterior distribution via Bayesian imputation. Specifically, we use the observed data to impute the unobserved potential outcomes from their posterior distribution. After imputing the missing potential outcomes, we can perform either between-unit or within-unit comparisons between treatment and control groups using any chosen discrepancy metric. Therefore, following the discussion in Section~\ref{subsec:limitations}, we focus on the following quantities:
\begin{equation}
  \label{eq:V_definitions}
  \begin{aligned}
    V_{Y} & = \Prob(Y_{1i} > Y_{0i}), \quad V_{S} & = \Prob(S_{1i} > S_{0i}), \quad \theta & = V_{Y} - V_{S}.
  \end{aligned}
\end{equation}
As is typically done in Bayesian analysis, we perform inference on \(\theta\) by sampling from its posterior distribution. At each iteration of our sampling algorithm, we impute the unobserved \(Y_{gi}\) and \(S_{gi}\) values (e.g., \(Y_{0i}\) and \(S_{0i}\) for treated units) from their posterior distribution. Using both the observed and imputed outcomes, we compute the probabilities in Equation~\eqref{eq:V_definitions} via Monte Carlo estimators, and obtain a posterior sample for \(\theta\) by taking their difference. Then, using posterior draws of \(\theta\), we can perform inference and assess whether \(S\) is a valid surrogate. Specifically, we compute a one-sided credible interval for \(\theta\) using a quantile of the \(T\) posterior draws, after discarding the first \(b\) iterations as burn-in:
\begin{equation}
    \widehat{\theta}_{1 - \alpha} 
    = \operatorname{quantile}\!\left( \widehat{\theta}^{(b+1:T)},\, 1 - \alpha \right),
\end{equation}
and perform the hypothesis test:
\begin{equation}
    \label{eq:hypothesis_test_theta}
    \begin{cases}
        H_{0}: \theta \geq \eta, \\
        H_{1}: \theta < \eta.
    \end{cases}
\end{equation}
where \(\eta\) is a pre-specified threshold for surrogate validation. We reject the null hypothesis if \(\widehat{\theta}_{1 - \alpha} < \eta\); otherwise, we fail to reject it. The full procedure is summarized in Algorithm~\ref{alg:bayesian_approach}.

The primary challenge of this methodology lies in sampling the unobserved potential outcomes. Clearly, for each unit \(i\), the sampling strategy to draw the missing potential outcomes from their posterior distribution crucially depends on the assumed data-generating mechanism, denoted as \(F_{i}\) in Equation~\eqref{eq:data_generating_process}. Therefore, in the next section, we propose two flexible parametric models that can be used for this sampling step.

\subsection{Parametric Models for the Data Generating Mechanism}\label{subsection:parametric_models}

The first model we consider excludes covariates, and therefore more closely aligns with the setting of Parast et al.\ \citep{parast2024rank}. Specifically, we define \(F_{i}\) from Equation~\eqref{eq:data_generating_process} as follows:
\begin{equation}
  \label{eq:parametric_model}
  \begin{aligned}
    P_{i} & \iid \; \mathcal{N}_{4}\!\left( \mu , \Sigma \right), \quad i=1,\ldots,n, \\
    \mu & \sim \; \mathcal{N}_{4}(\mu_{0}, \Sigma_{0}), \\
    \Sigma &= \text{diag}(\sigma_{1:4}) \,\Omega\, \text{diag}(\sigma_{1:4}), \\
    \sigma_{k} &\iid \; \text{Half-Normal}(0, s), \quad k = 1,2,3,4, \\
    \Omega & \sim \; \text{LKJ}(\tau),
  \end{aligned}
\end{equation}
where we place a Gaussian prior on the mean vector \(\mu\) and we adopt a separation strategy \citep{barnard2000modeling} for the covariance matrix \(\Sigma\), decomposing it into a correlation matrix \(\Omega\) and a scale vector \(\sigma_{1:4}\). This allows us to specify easily interpretable priors: an LKJ prior \citep{lewandowski2009generating} for the correlation and a half-normal prior for the scale.

However, one of the main advantages of our proposed Bayesian framework is its ability to naturally incorporate baseline covariates into the data-generating mechanism. To illustrate this feature, we consider a second model that introduces covariates through a linear regression structure, specifying \(F_{i}\) from Equation~\eqref{eq:data_generating_process} as follows:
\begin{equation}
  \label{eq:parametric_model_covariates}
  \begin{aligned}
    P_{i} &\overset{\text{ind.}}{\sim}\;
            \mathcal{N}_{4}\!\left( B X_{i} , \Sigma \right), \quad i=1,\ldots,n, \\
    B &= \big[ \beta_{1} \; \beta_{2} \; \beta_{3} \; \beta_{4} \big]^{\top}, \\
    \beta_{k} & \iid \; \mathcal{N}_{d}(\mu_{\beta}, \Sigma_{\beta}), \quad k = 1,2,3,4, \\
    \Sigma &= \text{diag}(\sigma_{1:4}) \,\Omega\, \text{diag}(\sigma_{1:4}), \\
    \sigma_{k} &\iid \; \text{Half-Normal}(0, s), \quad k = 1,2,3,4, \\
    \Omega & \sim \; \text{LKJ}(\tau),
  \end{aligned}
\end{equation}
where we place a Gaussian prior on the regression coefficients in the matrix \(B\) and we adopt the same separation strategy for the covariance matrix \(\Sigma\) as in the first model.

\subsection{Identifiability and Interpretation of the Posterior}
\label{subsection:identifiability}

Both formulations in Equations~\eqref{eq:parametric_model} and \eqref{eq:parametric_model_covariates} are quite simple and intuitive. However, they are both plagued by a common non-identifiability issue. Specifically, it is a well-known fact in the causal inference literature that the joint distribution of the potential outcomes is not identifiable from the observed data, since only one potential outcome is observed for each unit. This issue is typically referred to as the ``fundamental problem of causal inference'' \citep{holland1986statistics}. Therefore, to understand how this issue manifests in our setting, it is useful to take a closer look at the correlation matrix \(\Omega\) in the two models:
\begin{align}
\label{eq:correlation_matrix}
  \Omega = \begin{bmatrix}
    1 & \textcolor{ForestGreen}{\rho_{Y_{1}S_{1}}} & \textcolor{BrickRed}{\rho_{Y_{1}Y_{0}}} & \textcolor{BurntOrange}{\rho_{Y_{1}S_{0}}} \\
    \textcolor{ForestGreen}{\rho_{Y_{1}S_{1}}} & 1 & \textcolor{BurntOrange}{\rho_{S_{1}Y_{0}}} & \textcolor{BrickRed}{\rho_{S_{1}S_{0}}} \\
    \textcolor{BrickRed}{\rho_{Y_{1}Y_{0}}} & \textcolor{BurntOrange}{\rho_{S_{1}Y_{0}}} & 1 & \textcolor{ForestGreen}{\rho_{Y_{0}S_{0}}} \\
    \textcolor{BurntOrange}{\rho_{Y_{1}S_{0}}}& \textcolor{BrickRed}{\rho_{S_{1}S_{{0}}}} & \textcolor{ForestGreen}{\rho_{Y_{0}S_{0}}}& 1
    \end{bmatrix}.
\end{align}
It is easy to see that the green entries in the correlation matrix from Equation~\eqref{eq:correlation_matrix} are identifiable from the data, since they involve correlations between potential outcomes that are observed together in the data. In contrast, the remaining entries are not identifiable from the data, since they involve correlations between potential outcomes that are never observed together for any unit. Moreover, as we showed in Equation~\eqref{eq:decomposition_theta} for the Gaussian case, the red entries directly affect the computation of \(\theta\), while the orange entries do not.

In principle, one could try to address this non-identifiability issue by imposing further assumptions on the correlation matrix \(\Omega\). For example, one could assume that all the non-identifiable entries are equal to zero, which would imply that the potential outcomes are independent of each other. However, this assumption is quite strong and is likely unreasonable in many applications. But if we are not willing to make such strong assumptions, how can we still make sense of the posterior distribution of \(\theta\) obtained from our Bayesian imputation approach? The key insight is that, even though the posterior distribution of \(\theta\) is not fully identified from the data, it still contains useful information about the parameter of interest. 

To understand this point, it is useful to consider the asymptotic behavior of the posterior distribution of \(\theta\) in terms of concentration and, ideally, consistency. In particular, it is a well-known result in the Bayesian literature that, under suitable regularity conditions, the posterior distribution of a parameter concentrates around the true value of the parameter as the sample size increases. To formalize this intuition, let us be more precise about what we mean by regularity for the purpose of Bayesian posterior consitency:
\begin{definition}[Regularity of a Bayesian model]
\label{def:regularity}
  Let the Bayesian model be defined as
  \begin{align}
    \begin{split}
      D_{i} \mid \phi & \iid f_{\phi}, \quad i = 1, \ldots, n, \\
      \phi & \sim \pi,
    \end{split}
  \end{align}
  where \(D_{i}\) is the observed data for unit \(i\), \(f_{\phi}\) is the likelihood function parametrized by \(\phi\), and \(\pi\) is a prior distribution over the parameter \(\phi\). We say that the Bayesian model is regular if the following conditions hold:
  \begin{itemize}
    \item \textbf{Kullback-Leibler support of the prior}: The prior \(\pi\) assigns positive mass to every Kullback-Leibler neighborhood of the true parameter \(\phi^{*}\), i.e., for every \(\varepsilon > 0\), we have that
    \begin{equation}
      \pi \left( \left\{ \phi : D_{\mathrm{KL}}(f_{\phi^{*}} \| f_{\phi}) < \varepsilon \right\} \right) > 0,
    \end{equation}
    where \(D_{\mathrm{KL}}(f_{\phi^{*}} \| f_{\phi})\) is the Kullback-Leibler divergence between the true distribution \(f_{\phi^{*}}\) and the model distribution \(f_{\phi}\).

    \item \textbf{Identifiability}: The model is identifiable, i.e., for every \(\phi \neq \phi'\), we have that \(f_{\phi} \neq f_{\phi'}\).
  \end{itemize}
\end{definition}
Then, given the above definition of regularity, we can consider the following result which is commonly known as Schwartz's theorem \citep{schwartz1965bayes}:
\begin{theorem}[Schwartz's theorem]
  Let the true data-generating process be defined as
  \begin{equation}
    D_{i} \iid f_{\phi^{*}}, \quad i = 1, \ldots, n,
  \end{equation}
  where \(D_{i}\) is the observed data for unit \(i\) and \(f_{\phi^{*}}\) is the true distribution of the data, parametrized by \(\phi^{*}\). Let the Bayesian model be defined as
  \begin{align}
    \begin{split}
      D_{i} \mid \phi & \iid f_{\phi}, \quad i = 1, \ldots, n, \\
      \phi & \sim \pi,
    \end{split}
  \end{align}
  where \(\pi\) is a prior distribution over the parameter \(\phi\). Then, if the Bayesian model is regular according to Definition~\ref{def:regularity}, the posterior distribution of \(\phi\) concentrates around the true parameter \(\phi^{*}\) as the sample size goes to infinity:
  \begin{equation}
    \pi \left( U^{c} \mid D_{1:n} \right) \overset{P_{\phi^{*}}}{\to} 0,
  \end{equation}
  for every neighborhood \(U\) of \(\phi^{*}\), where \(P_{\phi^{*}}\) is the probability measure induced by the true distribution \(f_{\phi^{*}}\).
\end{theorem}
The first condition of Schwartz's theorem is a relatively mild condition that can be easily satisfied by choosing reasonable priors, such as the ones we proposed in Section~\ref{subsection:parametric_models}. However, the second condition of Schwartz's theorem is not satisfied in our setting due to the non-identifiability issue discussed above. Therefore, we cannot guarantee that the posterior distribution of \(\theta\) will concentrate around the true value of \(\theta\) as the sample size increases. But then the question is: what can we say about the posterior distribution of \(\theta\) in this case? The key insight is that, even though the posterior distribution of \(\theta\) is not fully identified from the data, it still contains useful information about the parameter of interest. In particular, as the sample size increases, the posterior distribution of \(\theta\) will concentrate around a set of values that are compatible with the observed data and the model assumptions. This intuition can be formalized considering this formulation found in \cite{gustafson2010bayesian}:
\begin{theorem}
  Let the true data-generating process be defined as
  \begin{equation}
    D_{i} \iid f_{\phi^{*}}, \quad i = 1, \ldots, n,
  \end{equation}
  where \(D_{i}\) is the observed data for unit \(i\) and \(f_{\phi^{*}}\) is the true distribution of the data, parametrized by \(\phi^{*}\). Let the parameter \(\phi\) be decomposed as follows:
  \begin{equation}
    \phi = (\phi_{I}, \phi_{N}),
  \end{equation}
  where \(\phi_{I}\) is the identifiable part of the parameter and \(\phi_{N}\) is the non-identifiable part of the parameter. Let the Bayesian model be defined as
  \begin{align}
    \begin{split}
      D_{i} \mid \phi & \iid f_{\phi_{I}}, \quad i = 1, \ldots, n, \\
      \phi & \sim \pi,
    \end{split}
  \end{align}
  where \(\pi\) is a prior distribution over the parameter \(\phi\). Then, if the identifiable component of the Bayesian model is regular according to Definition~\ref{def:regularity}, the posterior distribution of \(\phi\) concentrates around the set of values that are compatible with the observed data and the model assumptions as the sample size goes to infinity:
  \begin{equation}
    \pi \left( \phi_{I} \in U_{I}, \phi_{N} \in U_{N} \mid D_{1:n} \right) \overset{P_{\phi^{*}}}{\to} 1,
  \end{equation}
  for every neighborhood \(U_{I}\) of the identifiable component \(\phi_{I}^{*}\) and every neighborhood \(U_{N}\) of the non-identifiable component \(\phi_{N}^{*}\), where \(P_{\phi^{*}}\) is the probability measure induced by the true distribution \(f_{\phi^{*}}\). Whereas, the posterior distribution of the non-identifiable component \(\phi_{N}\) will behave as follows:
  \begin{equation}
    \pi \left( \phi_{N} \in U_{N} \mid D_{1:n} \right) \overset{P_{\phi^{*}}}{\to} \pi \left( \phi_{N} \in U_{N} \mid \phi_{I} = \phi_{I}^{*} \right),
  \end{equation}
  for every neighborhood \(U_{N}\) of the non-identifiable component \(\phi_{N}^{*}\), where \(\pi \left( \phi_{N} \in U_{N} \mid \phi_{I} = \phi_{I}^{*} \right)\) is the conditional prior distribution of the non-identifiable component given the identifiable component evaluated at its true value.
\end{theorem}
What this last result tells us is that, even though the posterior distribution of \(\theta\) is not fully identified from the data, there is still a process of Bayesian learning that takes place as the sample size increases, and the posterior distribution of \(\theta\) will concentrate around a set of values that are compatible with the observed data and the model assumptions. Therefore, it is still possible to meaningfully perform inference on \(\theta\) for the purpose of surrogate validation. However, it is important to keep in mind that due to the lack of concentration for the non-identifiable component of the parameter, there is a non-vanishing uncertainty in the posterior distribution of \(\theta\) that persists even in the asymptotic limit. This feature of the proposed Bayesian imputation approach may appear to constitute an inherent limitation. Nevertheless, we argue that the methodology retains merit for at least two reasons.

First, the identifiability issue is not unique to our proposed Bayesian imputation approach, but rather it is a fundamental issue that arises in any approach that relies on the observed data to make inference about the joint distribution of the potential outcomes. Therefore, any method that attempts to assess surrogate validity using the observed data will have to contend with this issue in some form. Then, the non-vanishing uncertainty in the posterior distribution of \(\theta\) can be seen as a reflection of the inherent uncertainty in the problem of surrogate validation. However, since the amount of uncertainty injected by the prior does not disappear even asymptotically, it becomes of crucial importance to choose reasonable priors that reflect our beliefs about the data-generating process. In this regard, given the prior specifications chosen in Equations~\eqref{eq:parametric_model} and \eqref{eq:parametric_model_covariates} and, more specifically, the choice of the prior for the correlation matrix \(\Omega\), unless we have strong prior beliefs about the correlation structure of the potential outcomes, it is reasonable to choose a weakly informative prior, such as an LKJ prior with \(\tau = 1\), which corresponds to a uniform distribution over the space of correlation matrices.

Second, even though our proposed Bayesian imputation approach does not showcase the desirable property of consistency, it still provides useful information about surrogate validity and, most importantly, it may still be preferable to alternative approaches that may not suffer by this non-identifiability issue but are unable to correctly capture the causal relationship between the surrogate and the primary outcome. For example, as we showed in Section~\ref{subsec:limitations}, the discrepancy parameter \(\delta\) used by Parast et al.\ \citep{parast2024rank} may fail to capture the causal relationship between \(Y\) and \(S\) even in very simple settings, while our approach targets a parameter \(\theta\) that is able to correctly reflect surrogate validity. Hence, this suggests that using our approach can lead to more accurate assessments of surrogate validity than the rank-based approach of Parast et al.\ \citep{parast2024rank} and especially in small-sample settings where the lack of asymptotic consistency may not be as problematic.

\subsection{Selection Procedure for the Validation Threshold}\label{subsec:alternative_threshold}

We now turn to the challenge of selecting the prespecified threshold used to assess surrogate validity, denoted \(\varepsilon\) by Parast et al.\ \citep{parast2024rank} and \(\eta\) in Equation~\eqref{eq:hypothesis_test_theta}. As noted by Parast et al.\ \citep{parast2024rank}, the distribution of \(\widehat{\delta}\) is not fully specified under the null in Equation~\eqref{eq:hypothesis_test_delta}, so choosing the threshold \(\varepsilon\) is not a trivial task. They suggest choosing it to ensure a desired power to detect a treatment effect on \(S\). Here, we outline an alternative procedure for selecting \(\eta\) in Equation~\eqref{eq:hypothesis_test_theta}, similar in spirit but better suited to our Bayesian framework, which relies on a Bayes factor, the standard tool for Bayesian hypothesis testing.

The intuition is that we aim to set \(\eta\) such that when $V_S$ is close to $V_{Y}$ i.e., within \(\eta\), $S$ is a valid replacement of $Y$. To do this, we aim to identify $v_s$, which is defined as the value of \(V_{S}\) that ensures a power of \((1 - \beta)100\%\) for the following hypothesis test:
\begin{equation}
    \label{eq:hypothesis_test_V_S}
    \begin{cases}
        H_{0}: V_{S} = \tfrac{1}{2}, \\
        H_{1}: V_{S} > \tfrac{1}{2}
    \end{cases}
\end{equation}
implemented using a Bayes factor. Let $v_Y$ denote the hypothesized treatment effect on $Y$. One reasonable way to choose $v_Y$ is by using the posterior mean of $V_Y$ obtained from our Bayesian imputation framework. Then $v_{Y} - v_{S}$ reflects how far we are willing to let $v_{S}$ be from $v_{Y}$, while still ensuring a \((1 - \beta)100\%\) power. To this end, the value of \(\eta\) is defined as:
\begin{equation}
    \eta = \max \{ v_{Y} - v_{S}, 0 \}.
\end{equation}
We now describe the details behind obtaining the value $v_s$. To compute a Bayes factor that can be used to conduct the hypothesis test described in Equation~\eqref{eq:hypothesis_test_V_S}, we first need to specify a Bayesian model taking \(V_{S}\) as its unknown parameter:
\begin{equation}
    \begin{aligned}
        n \, \widehat{V}_{S} \mid V_{S} & \sim \text{Binomial}(n, V_{S}), ~~\mbox{and}~~ V_{S} & \sim p(V_{S}),
    \end{aligned}
\end{equation}
\vspace{-1mm} 
where the prior for \(V_{S}\) is a Beta distribution renormalized on \( \left( \frac{1}{2}, 1 \right)\):
\begin{equation}
    p(V_{S}) = \frac{\text{Beta}(V_{S} \mid a, b)}{\int_{\frac{1}{2}}^{1} \text{Beta}(v \mid a, b) \, dv} \; \forall \; V_{S} \in \left(\tfrac{1}{2}, 1\right),
\end{equation}
with \(a, b > 0\) suitably chosen prior parameters, and \(\widehat{V}_{S} = \frac{1}{n} \sum_{i=1}^{n} \one(S_{1i} > S_{0i})\). Note that \(\widehat{V}_{S}\) cannot be computed directly from the observed data, since \(S_{1i}\) and \(S_{0i}\) are never jointly observed for any unit \(i\). However, as we shall see below, this is not an issue, because the procedure to select \(v_{s}\) only requires the sampling distribution of \(\widehat{V}_{S}\) under the null hypothesis and alternative hypothesis, rather than an observed value for it.

Thus, the Bayes factor that is used to compare our competing hypotheses is:
\begin{equation}
    BF_{n} = \frac{m \left( \widehat{V}_{S} \mid H_{1} \right) }{m \left( \widehat{V}_{S} \mid H_{0} \right)},
\end{equation}
where \(m \left( \widehat{V}_{S} \mid H_{0} \right) \) and \(m \left( \widehat{V}_{S} \mid H_{1} \right) \) are the marginal likelihoods of \(\widehat{V}_{S}\) under the null and the alternative hypothesis, respectively. These quantities can be computed as follows:
\begin{eqnarray}
m \left(\widehat{V}_{S} \mid H_{0} \right) 
&=& \binom{n}{n \, \widehat{V}_{S}} \frac{1}{2^{n}} \\
m\left(\widehat{V}_{S} \mid H_1\right) 
&=& 
\binom{n}{n \, \widehat{V}_{S}} \frac{B_{\frac{1}{2}}^{1}\big(a + n \, \widehat{V}_{S}, \, b + n - n \, \widehat{V}_{S}\big)}{B_{\frac{1}{2}}^{1}(a, b)}
\end{eqnarray}
where \(B_{x_{0}}^{x_{1}}(a, b)\) is the incomplete Beta function with parameters \(a, b\) over the interval \([x_{0}, x_{1}]\). Therefore, we have that
\vspace{-2mm} 
\begin{equation}
    \begin{aligned}
      BF_{n} 
        & = \frac{B_{\frac{1}{2}}^{1}\big(a + n \, \widehat{V}_{S}, \, b + n - n \, \widehat{V}_{S}\big)}{B_{\frac{1}{2}}^{1}(a, b)} 2^n.
    \end{aligned}
\end{equation}
As usual, we would like to ensure a type I error probability of $\alpha$:
\begin{equation}
    \mathbb{P} (BF_{n} > BF_{n, \alpha} \mid H_0) = \alpha.
\end{equation}
The quantity \(BF_{n, \alpha}\) must be computed numerically, which is feasible since \(n \, \widehat{V}_{S}\) ranges between \(0\) and \(n\), and therefore \(BF_{n}\) can take at most \(n + 1\) distinct values. Thus, \(BF_{n, \alpha}\) can be obtained as the \((1 - \alpha)\)-quantile of \(BF_{n}\). Note that the distribution of \(BF_{n}\) under the null hypothesis can be derived using the following formula:
\begin{equation}
    \mathbb{P} (BF_{n} = x_{k} \mid H_0) = \binom{n}{k} \frac{1}{2^{n}} \text{ for } k = 0, \ldots, n,
    \label{eq:BF_distribution_1}
\end{equation}
where
\begin{equation}
    x_{k} = \frac{B_{\frac{1}{2}}^{1}\big(a + k, \, b + n - k \big)}{B_{\frac{1}{2}}^{1}(a, b)} 2^n \text{ for } k = 0, \ldots, n.
    \label{eq:BF_distribution_2}
\end{equation}
The value of $BF_{n, \alpha}$ thus depends on the sample size $n$, the probability of type I error \( \alpha \) and the prior parameters $a, b$. 
Given $BF_{n, \alpha}$, we can define $v_{S}$ as the value such that
\begin{equation}
    \mathbb{P} \left( BF_{n} > BF_{n, \alpha} \mid V_{S} = v_{S} > \frac{1}{2} \right) = 1 - \beta.
    \label{eq:V_S_power}
\end{equation}
 It useful to observe that \(\forall \, v_{S} \in \left(\frac{1}{2}, 1\right)\),
\begin{equation}
v_{S} \mapsto \mathbb{P} \big( BF_{n} > BF_{n, \alpha} \mid V_S = v_{S} \big)
\end{equation}
is a monotone function. Therefore, there exists a unique \(v_{S}\) satisfying Equation~\eqref{eq:V_S_power}. The probability can be evaluated using the same procedure as in Equations~\eqref{eq:BF_distribution_1} and~\eqref{eq:BF_distribution_2} for the distribution of \(BF_{n}\) given \(V_S = v_{S}\). Since the function is continuous and monotone, \(v_{S}\) can be obtained using standard numerical root-finding algorithms, such as \texttt{uniroot} in \texttt{R}.

\section{Simulation Study}\label{sec:simulation_study}

We evaluated the performance of our proposed Bayesian imputation-based method against the frequentist rank-based approach of Parast et al.\ \citep{parast2024rank} in terms of test coverage and power via a simulation study in five settings.

Specifically, in Settings 1 to 4, we replicated the data-generating mechanisms used by Parast et al.\ \citep{parast2024rank} to facilitate direct comparison. In Settings 1 to 3 the potential outcomes for the primary outcome were generated as Gaussian variables, but in setting 1 the surrogate was designed to be useless, since its potential outcomes were generated independently of the primary outcome, while in settings 2 and 3 the potential outcomes for the surrogate were obtained by applying a linear transformation of the primary outcome, with the addition of Gaussian noise. Moreover, in Setting 2, the surrogate was designed to be perfect, since the noise was negligible, while in Setting 3, the surrogate was designed to be imperfect, since the noise was more substantial. In contrast, in Setting 4 the outcome was generated from a non-Gaussian distribution and the surrogate was generated by applying a non-linear transformation with the addition of non-Gaussian noise. In all of these settings, we implemented the model in Equation~\eqref{eq:parametric_model} without covariates. Given the true data generating mechanism, one would expect our proposed method to work well in Settings 1-3, but possibly not work well in Setting 4.

In Setting 5, we purposefully aimed to break the frequentist rank-based method and highlight the advantage of our proposed approach based on the model in Equation~\eqref{eq:parametric_model_covariates} with covariates. In particular, in Setting 5, we generated a baseline covariate as  \(X_{i} \iid \text{Bernoulli}(0.5)\), and the distribution of \(P_{i}\) given \(X_{i}\) was defined as follows:
\begin{equation}
  P_{i} \mid X_{i} \; \overset{\text{\footnotesize ind.}}{\sim} \; X_{i} \; \mathcal{N}_{4} \left(\mu^{1}, \Sigma^{1}\right)
  + (1 - X_{i}) \; \mathcal{N}_{4}\left(\mu^{0}, \Sigma^{0}\right), 
  \quad i = 1,\ldots,n,
\end{equation}
with the following specific parameter values:
\begin{align}
    \mu^{0} &= (5, 5, 0, 0)^\top, &
    \mu^{1} &= (5, -5, 0, -10)^\top, \notag\\
    \Sigma^{0} &= \Sigma^{1} = 
    \begin{bmatrix}
      A & 0 \\
      0 & A
    \end{bmatrix}, &
    A &= \begin{bmatrix} 1 & 1 \\ 1 & 2 \end{bmatrix}.
\end{align}
This construction implies that, conditionally on the covariates, the potential outcomes are not identically distributed across units, violating a key assumption of the Mann-Whitney test. More specifically, we chose the generating parameters to maximize the discrepancy between \(\delta\) and \(\theta\), which are equal to \(0.25\) and \(0\), respectively, while keeping the generating process as regular as possible. In fact, it is easy to verify that in this setting the surrogate is just a replica of the primary outcome with an addition of unitary Gaussian noise, with treatment effect homogeneity across the two subpopulations defined by \(X\) and uncorrelated potential outcomes. Therefore, we would hope that any surrogate validation method would correctly identify $S$ as a valid surrogate. 

We ran \(500\) simulations for each setting, with a sample size of \(n = 50\) overall and a significance level of \(\alpha = 0.05\). For our Bayesian method, we ran \(T = 500\) posterior sampling iterations with a burn-in period of \(b = 125\). We set the prior parameters across the two models as follows: \(\mu_{0} = (0, 0, 0, 0)^{\top}\), \(\Sigma_{0} = 10 \, I_{4}\), \(s = 2\), \(\tau = 1\),  \(\mu_{\beta} = (0, 0, 0, 0)^{\top}\), \(\Sigma_{\beta} = 10 \, I_{4}\). To compute the surrogate validation threshold \(\eta\), we set \(\beta = 0.2\), and a uniform prior for \(V_{S}\) by setting \(a = b = 1\). We quantify performance in terms of coverage of the confidence/credible intervals of the true parameter (\(\delta\) for the frequentist method and \(\theta\) for the Bayesian method) and power, by which we mean the proportion of simulation iterations where the method deemed the surrogate to be valid. In Setting $1$, the surrogate is not valid, and thus, this is the Type 1 error rate. 

Results are shown in Table~\ref{tab:coverage_power_comparison}. In Settings \(1\) and \(2\), the rank-based method of Parast et al.\ \citep{parast2024rank} and the Bayesian imputation-based method perform similarly. In in Setting \(3\),improved performance is observed for the Bayesian method with higher coverage and higher power. As expected, in Setting \(4\), where the data-generating process is non-Gaussian and non-linear, the nonparametric rank-based method outperforms the Bayesian method in terms of power. Setting  \(5\) highlights the true strength of our proposed Bayesian imputation-based method, where the frequentist method very rarely identifies the surrogate as valid, while our method correctly recognizes the surrogate as valid in all iterations. This is not surprising, since the data-generating process in Setting \(5\) was purposely designed to illustrate the limitations of the  the rank-based method.  The proposed Bayesian method is able to adjust for the binary covariate in this setting, and correctly capture surrogacy. The results demonstrate the power gains of our proposed methodology, though we also intentionally explore its limitations under model misspecification.

\section{Discussion}\label{sec:discussion}

We proposed a Bayesian imputation framework for surrogate evaluation in randomized clinical trials. Our method overcomes limitations of the frequentist rank-based approach, including lack of causal interpretation, inability to use baseline covariates, and low power, particularly in  in covariate-dependent settings. By imputing unobserved potential outcomes, it enables flexible treatment-control comparisons and inference on surrogate-primary associations. Incorporating baseline covariates improves applicability to complex scenarios, as shown in simulations where our method achieved perfect power versus \(0.075\) for the existing approach.

While our approach overcomes some limitations of the rank-based approach, there is room for further improvement, which we are actively pursuing. A key limitation is the reliance on parametric assumptions about the data-generating mechanism, which are vulnerable to model mis-specification. One possible direction, which we plan for future work, is to relax these assumptions by extending the framework to a Bayesian nonparametric setting \citep{ghosal2017fundamentals}, allowing a data-driven characterization of both the surrogate-outcome relationship and the covariate-outcome association. However, this flexibility comes at a cost of increased computational complexity and larger required sample sizes \citep{miller2018mixture}. Finally, sensitivity analyses examining robustness of our conclusions 
to unidentified covariance parameters would also be useful, and we aim to incorporate such results into our framework and future software.

\bibliography{Surrogate_bib.bib}

\vspace{3em}



\clearpage

\begin{longtable}[]{@{}cccccc@{}}
\caption{Simulation results comparing the frequentist rank-based method and the proposed 
Bayesian imputation-based method in terms of coverage and power in Settings 1--5, summarized 
over 500 replications with $n=50$ per replication. Bolding indicates the better performing method.}
\label{tab:coverage_power_comparison} \\
\toprule
\textbf{Setting} & \textbf{Valid Surrogate} & \multicolumn{2}{c}{\textbf{Coverage}} & \multicolumn{2}{c}{\textbf{Power}} \\
\cmidrule(lr){3-4} \cmidrule(lr){5-6}
& & Frequentist & Bayesian & Frequentist & Bayesian \\
\midrule
\endfirsthead
\multicolumn{6}{l}{\textit{(continued)}} \\
\toprule
\textbf{Setting} & \textbf{Valid Surrogate} & \multicolumn{2}{c}{\textbf{Coverage}} & \multicolumn{2}{c}{\textbf{Power}} \\
\cmidrule(lr){3-4} \cmidrule(lr){5-6}
& & Frequentist & Bayesian & Frequentist & Bayesian \\
\midrule
\endhead
\midrule
\multicolumn{6}{r}{\textit{(continued on next page)}} \\
\endfoot
\bottomrule
\endlastfoot
1  & No & 0.98 & \textbf{1.00} & 0.00 & 0.00 \\
2  & Yes & 0.98 & \textbf{1.00} & 1.00 & 1.00 \\
3 & Yes & 0.90 & \textbf{0.98} & 0.54 & \textbf{0.64} \\
4 (Misspecified) & Yes & 0.96 & \textbf{1.00} & \textbf{0.56} & 0.06 \\
5 (Binary Covariate) & Yes & 0.925 & \textbf{0.975} & 0.075 & \textbf{1.00} \\
\end{longtable}

\vspace{3em}
\clearpage

\begin{algorithm}[htbp!]
\caption{Bayesian Imputation-Based Approach.}
\label{alg:bayesian_approach}
\begin{algorithmic}[1]
\Require Observed data \(\{Y_i, S_i, Z_{i}, X_{i}\}_{i=1}^n\), prior parameters, number of posterior iterations \(T\), burn-in \(b\), surrogate threshold \(\eta\), significance level \(\alpha\)
\For{\(t = 1\) to \(T\)} 
    \For{\(i = 1\) to \(n\)}
        \If{\(Z_{i} = 1\)} 
            \State Sample \(P^{(t)}_{0i} \sim p \left( P^{(t)}_{0i} \mid P_{1i}, X_{i}, Z_{i}=1\right)\)
        \Else
            \State Sample \(P^{(t)}_{1i} \sim p \left( P^{(t)}_{1i} \mid P_{0i}, X_{i}, Z_{i}=0\right)\)
        \EndIf
    \EndFor
    \State \(\widehat{V}_Y^{(t)} \gets \frac{1}{n} \sum_{i=1}^{n} \one \big(Y_{1i}^{(t)} > Y_{0i}^{(t)}\big)\)
    \State \(\widehat{V}_{S}^{(t)} \gets \frac{1}{n} \sum_{i=1}^{n} \one \big(S_{1i}^{(t)} > S_{0i}^{(t)}\big)\)
    \State \(\widehat{\theta}^{(t)} \gets \widehat{V}_Y^{(t)} - \widehat{V}_{S}^{(t)}\)
\EndFor
\State \(\widehat{\theta}_{1-\alpha} \gets \operatorname{quantile}\!\Big(\widehat{\theta}^{(b+1:T)}, 1-\alpha\Big)\)
\If{\(\widehat{\theta}_{1-\alpha} < \eta\)}
    \State Reject \(H_0\)
\Else
    \State Fail to reject \(H_0\)
\EndIf
\end{algorithmic}
\end{algorithm}

\end{document}